\documentclass[12pt]{article}
\usepackage{graphicx}
\usepackage{amssymb,amsmath,amsfonts,palatino,amsthm}
\usepackage{amssymb}
\usepackage{epstopdf}
\usepackage{accents} 
\usepackage{nicefrac} 
\newcommand{\beq}{\begin{equation}}
\newcommand{\eeq}{\end{equation}}

\usepackage{bbm}

\def \al{\alpha}

\def \ga{\gamma}
\def \de{\delta}
\def \ep{\epsilon}

\def \io{\iota}
\def \ka{\kappa}

\def \rh{\rho}
\def \si{\sigma}

\def \ph{\phi}
\def \ch{\chi}
\def \ps{\psi}
\def \om{\omega}

\def \La{\Lambda}
\def \Si{\Sigma}

\def \Ph{\Phi}

\def \pa{\partial}

\def \lb{\left[}
\def \rb{\right]}
\def \lp{\left(}
\def \rp{\right)}

\def \p#1{\phantom{#1}}

\def \fr#1#2{{\textstyle \frac{#1}{#2}}}
\def \ha{\fr{1}{2}}

\def \f#1{\underaccent{\!\p{.}\! \raisebox{0pt}{\_} }{{#1}}}
\def \ff#1{\underaccent{=}{{#1}}}

\def \nf#1{\underaccent{\!\p{.} \sim}{{#1}}}
\def \ve#1{\accentset{\rightharpoonup}{{#1}}}

\newcommand{\mfr}{\mathfrak}

\begin{document}

\title{Unification of gravity, gauge fields, \\ and Higgs bosons}

\author{A. Garrett Lisi$^a$,\, Lee Smolin$^b$, and Simone Speziale$^c$  \\
\small{$^a$\emph{95 Miner Pl, Makawao, HI 96768}}\\
\small{$^b$\emph{Perimeter Institute for Theoretical Physics, 31 Caroline Street North, Waterloo, Ontario N2J 2Y5, Canada}} \\
\small{$^c$\emph{Centre de Physique Th\'eorique de Luminy, Case 907, 13288, Marseille, France}} 
}
\date{\today}
\maketitle
\begin{abstract}

We consider a diffeomorphism invariant theory of a gauge field valued in a Lie algebra that breaks spontaneously to the direct sum of the spacetime Lorentz algebra, a Yang-Mills algebra, and their complement. Beginning with a fully gauge invariant action -- an extension of the Plebanski action for general relativity -- we recover the action for gravity, Yang-Mills, and Higgs fields. The low-energy coupling constants, obtained after symmetry breaking, are all functions of the single parameter present in the initial action and the vacuum expectation value of the Higgs.

\end{abstract}

\section{Introduction}
  
While general relativity was originally formulated by Einstein as the dynamics of a metric, $g_{\mu \nu}$, it has 
been known since the work of Plebanski, Ashtekar and others \cite{Plebanski,Ashtekar,Capovilla} 
that, at a deeper level, the true configuration variable of general relativity is a connection -- corresponding to the gauging of the local Lorentz group, $SO(1,3)$, the spin group, $Spin(1,3)$, or a chiral subgroup.  The metric is relegated to an auxiliary role, analogous to that of the Higgs field. In this more natural, geometric formulation, the fundamental mathematical structure describing spacetime is not a pseudo-Riemannian manifold, but a principal bundle with connection over a four dimensional base manifold. The primacy of the connection is further evidenced by the simplified form of the gravitational action and equations of motion, which are polynomial in the connection formulation but non-polynomial in the original metric formulation. Thus, the idea that forces are described by a gauge connection is common to both Yang-Mills theory and general relativity.  This suggests an approach to the unification of gravity with the other forces in which the fundamental variable is a connection, $\f{H}$, valued in a Lie algebra, $\mfr{g}$, that includes a subalgebra $\mfr{h} \equiv \mfr{g}^{spacetime} \oplus \mfr{g}^{YM}$ -- the direct sum of the Lorentz algebra (or a chiral subalgebra of it) and a Yang-Mills gauge algebra.  In this work we propose a fully $\mfr{g}$-invariant gauge theory that breaks spontaneously to $\mfr{h}$ and yields gravity coupled to Yang-Mills theory.

It is well known, from the no-go theorem of Coleman and Mandula \cite{Coleman}, that {\it when global symmetries of the $S$-matrix are concerned,} such a unification cannot be accomplished without supersymmetry.
However, this result does not contradict our unification program because a spacetime geometry that could be used to define the $S$-matrix only exists after the $\mfr{g}$ symmetry has broken down to the direct sum, $\mfr{h}$. Before symmetry breaking, there is no metric and thus no $S$-matrix -- a loophole allowing the unification of gravity and gauge fields \cite{Percacci}.

In previous work we have taken steps towards such a formulation. In \cite{garrett-E8} one of us proposed a unification of this type, based on an extension of a formulation of general relativity described originally by MacDowell and Mansouri \cite{MacDowell}. 
In particular, the unification includes the {\it frame-Higgs} field for the breaking of $\mfr{g}$ in an elegant way, arising naturally as the non-$\mfr{h}$ components of the connection, $\f{H}$. The group there considered is $E8$,
but several of the results concerning the bosonic sector hold also for more general choices of $\mfr{g}$, as long as $\mfr{h}$ is contained. A weakness of this previous work was that the action in \cite{garrett-E8} was not fully $\mfr{g}$ gauge invariant, but only invariant under $\mfr{h}$. 

Subsequently \cite{ls-extended}, one of us proposed the formulation of a fully $\mfr{g}$-invariant theory via an extension of the covariant Plebanski action \cite{Capovilla,Mike,De Pietri}.\footnote{This kind of extension had previously been studied in \cite{Krasnov} for the self-dual Plebanski action, where $\mfr{g}^{spacetime}=\mfr{su(2)}$ (see also \cite{Bengtsson,Freidel,Ishibashi}). Recently in \cite{kirill-latest}, the same idea of grand unification put forward in \cite{ls-extended} and here developed, has been considered in the self-dual framework.}   
In this formulation there is a natural symmetry breaking mechanism that leads to a classical solution with unbroken gauge symmetry, $\mfr{h}$, and is characterized by a de Sitter background spacetime and approximate Yang-Mills dynamics. As anticipated in that paper, the complete spectrum of the theory also contains Higgs bosons.

In this paper, we show explicitly that this type of extended action can lead to gravity, Yang-Mills dynamics, and include the Higgs.
As in \cite{garrett-E8}, a frame-Higgs field for breaking $\mfr{g}$ naturally arises as the off-diagonal components of the connection in the solutions to the equations of motion. These off-diagonal components of the connection are one-forms
transforming as a vector under $\mfr{g}^{spacetime}$ (identified as the gravitational frame field) and in some representation space of $\mfr{g}^{YM}$ (identified with a Higgs multiplet). When $\mfr{g}^{YM}$ is also a spin group, these off-diagonal components factor into the gravitational frame times a zero-form valued in the vector representation space of $\mfr{g}^{YM}$ -- the Higgs field. The reader familiar with the Plebanski formalism might recognize this mechanism as a version of the simplicity constraints extracting the frame field from a two-form.

The coupling constants of the low-energy theory after symmetry breaking are Newton's constant, $G_N$, the cosmological constant, $\Lambda$, the Yang-Mills coupling constant, $g_{YM}$, and the parameters of the Higgs. These are all functions of the parameter present in the initial action.

\section{The action}

Our starting point for the unification of gravity and gauge fields is a polynomial action similar to the one proposed in \cite{ls-extended}. Here we focus on the bosonic sector, and discuss the coupling with fermions briefly in Section \ref{ferm}. 

We use $\mfr{g}^{spacetime} = spin(1,3)$ for the gravitational gauge algebra and presume the Yang-Mills gauge algebra to also be a spin algebra, $\mfr{g}^{YM} = spin(N)$, consistent with embedding the standard model gauge group in the $Spin(10)$ grand unified theory. Consequently, we presume the full initial gauge algebra to be $\mfr{g} = spin(1\!+\!N,3)$.
Following the discussion in the introduction, the unified bosonic field is a $spin(1\!+\!N,3)$ valued connection one-form over a four dimensional base manifold -- locally, $\f{H} = \f{dx^\mu} \ha H_\mu^{\p{\mu}IJ} \ga_{IJ}$. Here the bivector generators $\ga_{IJ}$ of $spin(1\!+\!N,3)$ have indices running over all $(4\!+\!N)\times(4\!+\!N)$ values,\footnote{This will make $\ha$ factors necessary in expressions involving redundant sums.} and can be understood as the product of $Cl(1\!+\!N,3)$ Clifford algebra basis vectors, $\ga_{IJ}=\ga_{I} \ga_{J}$.

The action we consider is the $\mfr{g}$-invariant action,
\begin{eqnarray}
S(H,B,\Phi) = \fr{1}{g} \int_{\cal M} \left< \ff{B} \ff{F} + \ff{B} \Phi \ff{B} + \fr{1}{3} \ff{B} \Phi \Phi \Phi \ff{B} \right>
\label{action1}
\end{eqnarray}
in which $\ff{F} = \f{d} \f{H} + \ha \lb \f{H}, \f{H} \rb$ is the curvature, $\ff{B} = \ha \ff{B}^{IJ} \ga_{IJ}$ is a $spin(1\!+\!N,3)$ valued 2-form field, and $\Phi$ is a symmetric linear operator which takes bivectors to bivectors and 2-forms to 2-forms. The wedge product is assumed between forms, which have underlines to designate their grade, and the angle brackets are shorthand for the trace -- equivalent to taking the Clifford scalar part. In this action, the connection, $\f{H}$, is the "physical" variable describing the geometry of the principal bundle, while $\ff{B}$ and $\Phi$ are auxiliary fields, with parameter $g$. 
Written out with indices, the second term in the action is
$$
\left< \ff{B} \Phi \ff{B} \right> = \nf{d^4 x} \, \fr{1}{32} \ep^{\mu \nu \rh \si} B_{\mu \nu IJ} \Phi_{\rh \si \p{\ph \ch IJ} KL}^{\p{\rh \si} \ph \ch IJ} B_{\ph \ch}^{\p{\ph \ch} KL}
$$
Varying $\f{H}$, $\ff{B}$ and $\Phi$, the field equations are:
\begin{eqnarray}
\f{\cal D} \ff{B} \!\! & \! = \! & \!\! \f{d} \ff{B} + \lb \f{H} , \ff{B} \rb  = 0
\label{eq2} \\
\ff{F} \!\!&\!=\!&\!\! -2 \lp \Phi + \fr{1}{3} \Phi \Phi \Phi \rp \ff{B}
\label{eq1} \\
\fr{1}{32} \ep^{\mu \nu \rh \si} B_{\mu \nu IJ} B_{\ph \ch}^{\p{\ph \ch} KL} \!\! & \! = \! & \!\!
- \fr{1}{512} \ep^{\mu \nu \rh \si} B_{\mu \nu IJ}
\Phi_{\ph \ch \p{\ps \om KL} MN}^{\p{\ph \ch} \ps \om KL}
\Phi_{\ps \om \p{\io \ka MN} PQ}^{\p{\ps \om} \io \ka MN}
B_{\io \ka}^{\p{\ph \ch} PQ}
\label{eq3}
\end{eqnarray}
The first equation (\ref{eq2}) may be thought of as describing the dynamics, while (\ref{eq1}) and (\ref{eq3}) determine $\ff{B}$ and $\Ph$. Note that (\ref{eq3}) does not necessarily constrain $\ff{B}$, and is satisfied by $\Ph$ provided
\beq
\ff{B} = - \Ph \Ph \ff{B}
\label{phieq}
\eeq

The specified action (\ref{action1}) is a modification of the action presented in \cite{ls-extended}, with $\Ph$ generalized to act on 2-forms as well as on Lie algebra generators. A specific "Mexican hat" potential for $\Ph$ in the action (\ref{action1}) has been chosen to allow symmetry breaking to a nontrivial vacuum expectation value (vev). We choose this specific action because it leads to a particularly simple analysis. However, more general actions, in which the $\Phi^3$ term is replaced by an arbitrary potential, $U(\Phi)$, can also be used as candidates for unification, and analyzed along similar lines as below.

\subsection{Symmetry breaking}

The action (\ref{action1}) and equations of motion are symmetric under $\mfr{g}$. We now exhibit a classical solution of the theory which distinguishes a $spin(1,3)$ subalgebra. A similar action and symmetry breaking mechanism has already been shown to lead to $spin(N)$ Yang-Mills theory on de Sitter space \cite{ls-extended}. Here we show a spontaneous symmetry breaking of our new action that produces the dynamics of gravity and a Higgs field as well as Yang-Mills.

The field $\Ph$ takes 2-forms to 2-forms and bivectors to bivectors. This leads us to consider the relation of $\Ph$ to a Hodge duality operator, $*$, which takes 2-forms to 2-forms over a four dimensional manifold, and the Lie algebra duality operator, $\star$, for $spin(1,3)$, that takes bivectors to bivectors. The Lie algebra dual of a $spin(1,3)$ bivector, $B = \ha B^{ab} \ga_{ab}$, is
$$
\star B = \ha B^{ab} ( \star \ga_{ab} ) = \ha B^{ab} ( \ha \ep_{ab}^{\p{ab}cd} \ga_{cd} ) = \ha B^{\star cd} \ga_{cd}
$$
with $\ep^{abcd}$ the permutation symbol and $B^{\star cd} = \ha B^{ab} \ep_{ab}^{\p{ab}cd}$. We use lower case letters early in the alphabet, $a,b,... \in \{1,2,3,4\}$, to sum over a subset of $I,J,... \in \{1,2,...,4\!+\!N\}$ and thereby choose a $spin(1,3)$ subalgebra of $spin(1\!+\!N,3)$. We will use later letters, $m,n,... \in \{5,6,...,4\!+\!N\}$, to sum over the rest. To build the Hodge star operator on 2-forms we must presume the existence of a nondegenerate gravitational frame, $\f{e}{}^a = \f{dx^\mu} (e_\mu)^a$, with a set of orthonormal basis vectors from its inverse, $\ve{e}_a = (e_a)^\mu \ve{\pa}_\mu$. Using this frame, the Hodge dual of any 2-form, $\ff{B} = \ha \f{e}^a \f{e}^b B_{ab}$, is
$$
* \ff{B} =  \ha ( * \f{e}^a \f{e}^b) B_{ab} =  \ha ( \ha \ep^{ab}_{\p{ab}cd} \f{e}^c \f{e}^d ) B_{ab} = \ha \f{e}^c \f{e}^d B_{* cd}
$$
with $B_{* cd} = \ha B_{ab} \ep^{ab}_{\p{ab}cd}$. If we define the frame as a Clifford vector valued 1-form, $\f{e}=\f{e}{}^a \ga_a$, we can define the gravitational area field,
$$
\ff{\Si} = \f{e} \f{e} = \f{e}{}^a \f{e}{}^{b} \ga_{ab} = \f{dx^\mu} \f{dx^\nu} (e_\mu)^a (e_\nu)^b \ga_{ab}
$$
a $spin(1,3)$ valued 2-form, with its Hodge dual the same as its $spin(1,3)$ dual,
\beq
* \ff{\Si} = \star \ff{\Si}
\label{ssi}
\eeq
Applying the Hodge dual or the $spin(1,3)$ dual twice gives the negative identity, $**=\star \star=-1$.

Using these two duality operators, which are symmetric linear operators taking bivectors to bivectors and 2-forms to 2-forms, we make a symmetry breaking ansatz for $\Ph$:
\beq
\Ph = a + \, b * + \, c \star + \, d * \star
\label{ansatz}
\eeq
in which $\{a,b,c,d\}$ are parameters to be determined by the equations of motion. Using indices, the spacetime components of this $\Ph$ are
$$
\Phi_{\mu \nu \p{\rh \si ab} cd }^{\p{\mu \nu} \rh \si ab}
= a \, \de_{\mu \nu}^{\rh \si} \de_{cd}^{ab} 
+ b \, (e_\mu)^e (e_\nu)^f \ep_{ef}^{\p{ef}gh} (e_g)^\rh (e_h)^\si \de_{cd}^{ab} 
+ c \, \de_{\mu \nu}^{\rh \si} \ep^{ab}_{\p{ab} cd} 
+ d \, \ep_{\mu \nu}^{\p{\mu \nu} \rh \si} \ep^{ab}_{\p{ab} cd}
$$
We next consider the validity of this ansatz for $\Ph$, and its implications for $\ff{B}$ and $\f{H}$, following from the equations of motion.

From our ansatz (\ref{ansatz}), we have
$$
\Ph \Ph = (a^2-b^2-c^2+d^2) + \, 2(ab-cd)* + \, 2(ac-bd) \star + \, 2(ad+bc) * \star
$$
Considering our equation of motion (\ref{eq3}), which is implied by (\ref{phieq}), we see two possible classes of solutions. In the first class of solutions, in which
$$
a=0 \;\;\;\;\;\; b = 1 \;\;\;\;\;\; c = 0 \;\;\;\;\;\; d=0
$$
we have $\Ph \Ph = -1$, and this equation of motion is satisfied for any $\ff{B}$, without restriction. In the second class of solutions, for any other nontrivial values of $a,b,c, \mbox{and } d$, this equation of motion implies a set of restrictions on $\ff{B}$. Although this second class of solutions is interesting to consider, we focus from here forward on the first class of solutions, in which we have
\beq
\Ph = *
\label{hodge}
\eeq
with the Hodge constructed using some gravitational frame, $\f{e}$. To further demonstrate the viability of our ansatz, we need to show that the other field equations can also be satisfied, and relate our use of the gravitational frame in the Hodge star in $\Ph$ to degrees of freedom in our physical variable, $\f{H}$.

Our ansatz (\ref{hodge}) for $\Ph$ ensures the field equation (\ref{eq3}) is satisfied for all $\ff{B}$. This ansatz also allows us to invert field equation (\ref{eq1}) and solve it for $\ff{B}$ in terms of the curvature,
\beq
\ff{B} = \fr{3}{4} * \! \ff{F}
\label{BfromF}
\eeq
Using this in (\ref{eq2}), we now have only one field equation, the Yang-Mills equation in curved spacetime,
\beq
0 = \f{D} * \! \ff{F} = \f{d} * \! \ff{F} + \big[ \f{H} , * \ff{F} \big]
\label{YM}
\eeq
Applying the Hodge star (now operating on 3-forms to give 1-forms), this equation is
\beq
0 = \ve{D} \ff{F} = \ve{\de} \ff{F} + \big[ \ve{H} , \ff{F} \big]
\label{YMc}
\eeq
in which the codifferential operator, lowering the grade of a form by 1, is defined to be  $\ve{\de} = * \f{d} *$ and we also define the \emph{vector} operator, $\ve{H}$, as
$$
\ve{H} \ff{F} = * \f{H} \! * \ff{F} =  H_d \eta^{ad} \ve{e}{}_a \f{e}{}^b \f{e}{}^c \ha F_{bc} = H_d \eta^{ad} \de_a^{[b} \f{e}{}^{c]} F_{bc} = H^a F_{ac} \f{e}{}^c
$$
For the Yang-Mills field equation, it is usual to consider solutions approximated by small perturbations around $\f{H}=0$. However, there is a different sector of solutions, in which $\f{H}$ has a nonzero vacuum expectation value related to the gravitational frame, $\f{e}$, in the $*$ of $\Ph$.

\subsubsection{Symmetry breaking to gravity}

Since our ansatz for $\Ph$ distinguishes a $spin(1,3)$ subalgebra of $\mfr{g} = spin(1\!+\!N,3)$, we are free to label the separate parts of our connection accordingly,
\beq
\f{H} = \ha \f{\om} + \fr{1}{4} \f{E} + \f{A}
\label{H}
\eeq
with the gravitational spin connection, $\f{\om} = \ha \f{\om}{}^{ab} \ga_{ab}$, valued in $spin(1,3)$, the gauge field, $\f{A} = \ha \f{A}{}^{mn} \ga_{mn}$, valued in $spin(N)$, and the {\it frame-Higgs}, $\f{E} = \f{E}{}^{am} \ga_{am}$, valued in the "off diagonal" complement, $4N$, of $\mfr{h}=spin(1,3)+spin(N)$ in $\mfr{g}$. To see how the $\f{e}$ in $*$ relates to the $\f{E}$ part of $\f{H}$, we first make an ansatz for $\f{E}$. We consider that $\f{E}$ may be a simple bivector,
\beq
\f{E} = \f{e'} \ph = \f{dx^\mu} (e'_\mu)^a \ph^m \ga_{am}
\label{E}
\eeq
the Clifford product of a 1-form frame, $\f{e'} = \f{dx^\mu} (e'_\mu)^a \ga_a$, transforming as a vector under $spin(1,3)$, and a scalar Higgs multiplet, $\ph=\ph^m \ga_m$, a vector under $spin(N)$. At this point, we have defined two frames, the $\f{e}$ in $*$ and the $\f{e'}$ in $\f{H}$, which are related through the field equation.

With our labeling (\ref{H}) for the parts of the connection, and including our ansatz (\ref{E}) for $\f{E}$, the curvature of $\f{H}$ is
\beq
\ff{F} = \f{d} \f{H} + \f{H} \f{H} = \ha \lp \ff{R} - \fr{1}{8} \ff{\Si'} \ph^2 \rp + \fr{1}{4} \lp \ff{T} \ph - \f{e'} \f{D} \ph \rp + \ff{F}{}_A
\label{curvature}
\eeq
in which $\ff{R} = \f{d} \f{\om} + \ha \f{\om} \f{\om}$ is the Riemann curvature 2-form, $\ff{\Si'} = \f{e'}\f{e'}$ is an area field, $\ph^2 = \ph \cdot \ph = \ph^m \ph_m$ is the squared magnitude of the Higgs, $\ff{T} = \f{d}\f{e'} + \ha [ \f{\om} , \f{e'} ]$ is torsion, $\f{D} \ph = \f{d} \ph + [ \f{A}, \ph ]$ is the covariant derivative of the Higgs, and $\ff{F}{}_A = \f{d} \f{A} + \f{A} \f{A}$ is the curvature of the gauge field. Using this curvature in our field equation (\ref{YMc}), the resulting equations of motion separate into $spin(1,3)$, $4N$, and $spin(N)$ parts:
\begin{eqnarray}
0 \!\! & \! = \! & \!\!
 \ha \ve{\de} \lp \ff{R} - \fr{1}{8} \ff{\Si'} \ph^2 \rp
+ \fr{1}{4} \big[ \ve{\om}, \ff{R} \big]
- \fr{1}{32} \big[ \ve{\om}, \ff{\Si'} \big] \ph^2
- \fr{1}{16} \big[ \ve{e'}, \ff{T} \big] \ph^2
+ \fr{1}{16} \big[ \f{e'} , \ve{e'} \big] \ph \cdot\! \f{D} \ph \label{fe1} \\
0 \!\! & \! = \! & \!\!
 \fr{1}{4} \ve{\de} \lp \ff{T} \ph - \f{e'} \f{D} \ph \rp
+ \fr{1}{8} \big[ \ve{\om}, \ff{T} \big] \ph
- \fr{1}{8} \big[ \ve{\om}, \f{e'} \big] \f{D} \ph
+ \fr{1}{8} \big[ \ve{e'}, \ff{R} \big] \ph
- \fr{3}{32} \f{e'} \ph^3
- \fr{1}{4} \big[ \ve{A} , \ph \big] \ff{T} \;\;\;\;\; \nonumber \\
\!\! & \!  \! & \!\!
- \fr{1}{4} \f{e'} \big[ \ve{A} , \f{D} \ph \big]
+ \fr{1}{4} \big[ \ve{e'} \ph, \ff{F}{}_A \big]  \;\;\;\;\; \label{fe2} \\
0 \!\! & \! = \! & \!\!
\ve{\de} \ff{F}{}_A
+ \big[ \ve{A}, \ff{F}{}_A \big]
+ \fr{1}{4} \big[ \ph , \f{D} \ph \big] \label{fe3}
\end{eqnarray}
These equations of motion, which followed from our symmetry breaking ansatz (\ref{hodge}), allow a wide range of solutions. These solutions include not only the dynamics of the gauge fields, $\f{A}$, but a subset describing the dynamics of gravity. This subset of solutions can be described by the ansatz that the gravitational frame in the Hodge star, $\f{e}$, is equal to the $\f{e'}$ part of the connection,
\beq
\f{e'} = \f{e}
\label{ee}
\eeq
This ansatz, restricting to a subset of solutions, results in a significant simplification of the equations of motion. Also, if we ignore fermionic matter, there is no source for torsion, which we can then take to vanish,
\beq
0 = \ff{T} = \f{d} \f{e} + \ha [\f{\om},\f{e}] \label{T0}
\eeq
determining $\f{\om}$ from $\f{e}$, and further simplifying the equations of motion. With these simplifications, the field equations are
\begin{eqnarray}
0 \!\! & \! = \! & \!\!
 \ve{\de} \ff{R}
+ \fr{1}{2} \big[ \ve{\om}, \ff{R} \big]
+  \fr{1}{8} \big[ \f{e} , \ve{e} \big] \ph \cdot \! \f{D} \ph \label{DR} \\
0 \!\! & \! = \! & \!\!
\fr{1}{2} \f{R} \ph
- \fr{3}{8} \f{e} \ph^3
-  \f{e} \ve{\de} \f{D} \ph
-  \f{e} \big[ \ve{A} , \f{D} \ph \big]
+ \big[ \ve{e} \ph, \ff{F}{}_A \big] \;\;\;\;\;\label{einstein} \\
0 \!\! & \! = \! & \!\!
\ve{\de} \ff{F}{}_A
+ \big[ \ve{A}, \ff{F}{}_A \big]
+ \fr{1}{4} \big[ \ph , \f{D} \ph \big] \label{cYM}
\end{eqnarray}
in which $\f{R} = [ \ve{e}, \ff{R} ]$ is the Ricci curvature, a $Cl(1,3)$ vector valued 1-form. These equations of motion produce dynamics for gravity provided $\ph \neq 0$. Operating on (\ref{einstein}) with $\ve{e}$, we get
$$
0 = \ha R \ph - \fr{3}{2} \ph^3 - 4 \ve{D} \f{D} \ph + \ve{e} \, \big[ \ve{e} \ph, \ff{F}{}_A \big]
$$
in which $R=\ve{e} \cdot \! \f{R}$ is the scalar curvature. Plugging this back in to the equation of motion (\ref{einstein}) and factoring out a $\ph$, we obtain Einstein's equation for our gravitational dynamics,
\beq
\f{G} = \f{R} - \ha \f{e} R = - \fr{3}{4} \f{e} \ph^2 - 2 \f{e}  \ph^- \ve{D} \f{D} \ph - \ph^- \big[ \ve{e} \ph, \ff{F}{}_A \big]
\label{eineq}
\eeq
Note that, since we must also satisfy (\ref{DR}), governing the propagation of gravitational waves, and since the right hand side of (\ref{eineq}) is not quadratic in $\ff{F}{}_A$, we are dealing with a modified form of gravity.

The implication from these considerations is straight forward. Our complete symmetry breaking ansatz is that the $\f{E}$ part of the connection (\ref{H}) is a simple bivector, $\f{e}\phi$, with components $E_\mu^{\p{\mu}am} = (e_\mu)^a \phi^m$, which acquires a nonzero vacuum expectation value; and $\Ph$ is equal to $*$, with this Hodge star built using the same frame, $\f{e}$. The dynamics of gravity, Yang-Mills, and Higgs then comes directly as the subset of solutions consistent with this ansatz in our equations of motion, or in the action.

\subsection{Gravitational, Yang-Mills and Higgs action}

With our ansatz (\ref{hodge}) for $\Ph$, the action (\ref{action1}) becomes
\beq
S(H,e,B) = \fr{1}{g} \int \left< \ff{B} \ff{F} + \fr{2}{3} \ff{B} * \! \ff{B} \right>
\label{action2}
\eeq
Solving the equation of motion (\ref{eq1}) for $\ff{B}$ in terms of $\ff{F}$ gives (\ref{BfromF}), and using this in the action gives the Yang-Mills action,
\beq
S(H,e) = \fr{3}{8 g} \int \big< \ff{F} * \! \ff{F} \big>
\label{action3}
\eeq
With our ansatz (\ref{H}) for the connection, including (\ref{E}) for $\f{E}$, the curvature of $\f{H}$ is given by (\ref{curvature}). Using this decomposition of the curvature, the action is
\beq
S(H,e) =
\fr{3}{8 g} \int \big< 
- \fr{\ph^2}{16} \ff{R} * \! \ff{\Si'} + \fr{\ph^4}{256} \ff{\Si'} * \! \ff{\Si'} + \fr{1}{4} \ff{R} * \! \ff{R}
 + \fr{1}{16} \lp \ff{T} \ph \!-\! \f{e'} \f{D} \ph \rp * \! \lp \ff{T} \ph \!-\! \f{e'} \f{D} \ph \rp + \ff{F}{}_A * \! \ff{F}{}_A \big> \;\;\;
\label{action4}
\eeq
Restricting to the gravitational sector, in which $\f{e'}=\f{e}$, and presuming vanishing torsion, this action is
\beq
S(e,\ph,A) =
\fr{3}{8 g} \int \nf{d^4x} |e| \big( 
 - \fr{1}{16} \ph^2 R + \fr{3}{32} \ph^4 + \fr{1}{16} R_{ab}^{\p{ab}cd} R^{ab}_{\p{ab}cd}
 - \fr{1}{2} D_a \ph^m D^a \ph_m - \fr{1}{4} F_{ab}^{\p{ab}mn} F^{ab}_{\p{ab}mn} \big) \;\;\;
\label{action5}
\eeq
The first term is the Einstein-Hilbert action for gravity, with Newton's constant equal to $G_N = \fr{128 g}{3 v^2}$, where $v^2=<\!\phi_0^2\!>$ is defined to be the magnitude of the square of the vacuum expectation value of the Higgs field. The second term is the cosmological constant, $\La = \fr{3}{4} v^2$, consistent with (\ref{eineq}). The third term -- a modification to standard gravitation -- is a Stephenson-Kilmister-Yang (SKY) term \cite{yangs}, related to a Gauss-Bonnet topological action. The fourth term is the kinetic action for the Higgs field. And the fifth term is the standard action for Yang-Mills gauge fields, with coupling constant $g^2_{YM} = \fr{2 g}{3}$.

The nontrivial vacuum solution to this action is de Sitter spacetime with a non-vanishing Higgs vacuum expectation value, $\ph_0$. Specifically, the standard Higgs potential in (\ref{action4}) has an extrema at $\ph_0^2 = \fr{1}{3} R_0$ corresponding to a de Sitter spacetime background solution,\footnote{A particularly nice explicit expression for a de Sitter frame is
$$
\f{e}{}_0 = \f{dt} \ga_4 + \f{d a^1} \al \cosh(\fr{t}{\al}) \ga_1 + \f{d a^2} \al \cosh(\fr{t}{\al}) \sin(a^1) \ga_2 
+ \f{d a^3} \al \cosh(\fr{t}{\al}) \sin(a^1) \sin(a^2) \ga_3
$$
with $\al=\fr{2}{v}$.
}
\beq
\ff{R}{}_0 = \fr{v^2}{8} \ff{\Si}{}_0
\;\;\;\;\;\;\;\;\;\;\;\;\;\;
\f{R}{}_0 = \fr{3 v^2}{4} \f{e}{}_0
\;\;\;\;\;\;\;\;\;\;\;\;\;\;
R_0 = 3 v^2 = 4 \La
\label{desitter}
\eeq
which implies vanishing $\mfr{g}$ curvature, $\ff{F}{}_0 = \ha \ff{R}{}_0 - \fr{1}{16} \ff{\Si}{}_0 \ph_0^2 = 0$, solving the field equations (\ref{YMc}) and strictly minimizing the action (\ref{action3}). Physically, from the geometry of a principal bundle with connection over a four dimensional base manifold, the symmetry breaks and the frame-Higgs part of the connection acquires a nonzero vev,
$$
\f{H}{}_0 = \fr{1}{4} \f{e}_0 \ph_0
$$
corresponding to a de Sitter spacetime and Higgs background -- endowing spacetime with geometry and particles with mass. Local dynamics then exist as fluctuations with respect to this cosmological background.

All physical constants -- including Newton's constant, the cosmological constant, the Yang-Mills coupling and Higgs parameters -- derive solely from $g$ and the Higgs vev. The relations obtained are clearly far from observed values, which might suggest that the model considered here is too simple to have phenomenological applications. However, before drawing such a pessimistic conclusion, we should note that these are bare parameters, so  $g^2_{YM}$ refers to the single coupling constant of the unified gauge group at the Planck scale, and similarly for the cosmological constant.  It is then not impossible that these equalities hold in the neighborhood of a fixed point that governs the asymptotic high energy behavior of the unified theory. The fact that the gravitational and Yang-Mills couplings are explicitly related is a sign that we are dealing with a genuine unification of gravity and Yang-Mills theory. 

\section{Fermions and unification} \label{ferm}

Before closing, we make some comments on the coupling to fermions and the prospect of incorporating a grand unified theory extending the standard model of particle physics. 

The coupling of the unified bosonic connection to fermions occurs in the covariant Dirac derivative,
\beq
\f{D} \ps = (\f{d} + \f{H}) \ps = (\f{d} + \fr{1}{4}  \f{\om}^{ab} \ga_{ab} + \fr{1}{4} \f{e}^a \ph^m \ga_{am} + \ha \f{A}^{mn} \ga_{mn} ) \ps
\eeq
in which $\ps$ lives in a spinorial representation space of $Spin(1\!+\!N,3)$. After symmetry breaking, when $\f{\om}$ is the gravitational spin connection, $\f{e}$ is the gravitational frame, $\ph$ is a Higgs multiplet, and $\f{A}$ is a Yang-Mills field, this covariant Dirac derivative gives the correct interactions between these fields and a multiplet of Dirac fermions in curved spacetime. Since the spin connection appears explicitly in this covariant derivative, we expect fermions to act as a source of torsion.

Looking ahead to the details of a unification incorporating the standard model, we note that 
the standard model gauge algebra is a subalgebra of the Pati-Salam GUT algebra, which is a subalgebra of the $spin(10)$ GUT algebra,
\beq
su(2)_L \oplus u(1)_Y \oplus su(3) \subset su(2)_L \oplus su(2)_R \oplus su(4) = spin(4) \oplus spin(6) \subset spin(10)
\eeq
When this is used in the context of our unified description, $\f{H}$ is in $spin(11,3)$, one generation of standard model Dirac fermions live in a $32 \mathbb{C}$ (or $64 \mathbb{R}$) positive chiral spinor rep, the $su(2)_L$ acts correctly on the left-chiral fermions in this rep, and the Higgs multiplet can acquire a vev, such as $\phi_0 = v \ga_5$, giving Dirac masses to these fermions.

For alternative ideas on the coupling to fermions in this type of grand unification scheme, see \cite{Capovilla,Percacci,garrett-E8,ls-extended,Nesti}.  Also, instead of attempting unification through an extended Plebanski formulation, it is possible to study the case where the connection is valued in a Lie algebra, $\mfr{g}$, that includes a subalgebra $\mfr{h} \equiv su(2) \oplus \mfr{g}^{YM}$  in which $su(2)$ is a subalgebra of the local Lorentz algebra -- as investigated in \cite{kirill-latest}. This kind of reduced extended Plebanski theory is known to be simpler than that considered here, as the gravitational sector has only the usual spin two degrees of freedom.  

\section{Discussion and conclusions}

We have seen that a fully $\mfr{g}$ invariant action (\ref{action1}) for a unified bosonic connection (\ref{H}) can produce symmetry breaking leading to gravitational, Yang-Mills, and Higgs actions for the different parts of $\f{H}$. The form of the action -- a perturbation to a topological BF action -- is almost as simple as possible; it consists of cubic and quintic terms, with only a single derivative appearing in just a single term. By examining the dynamics of our action (\ref{action1}) within the sector of solutions in which $\f{E}=\f{e}\ph$ and $\Ph=*$ using the same $\f{e}$, we find the dynamics of gravity coupled to Yang-Mills and Higgs fields, described via field equations (\ref{DR},\ref{einstein},\ref{cYM}) and via an action (\ref{action5}). To recapitulate, our argument is:
\begin{enumerate}
\item We begin with a $\mfr{g}$ invariant action (\ref{action1}).
\item We make two ansatze related to a symmetry breaking: (\ref{hodge}) for $\Phi$ and (\ref{E}) for part of $\f{H}$.
\item The resulting field equations are (\ref{fe1}-\ref{fe3}), which admit infinite solutions, and correspond to the action (\ref{action4}).
\item Among these solutions, we choose to look at those in which the frames are matched (\ref{ee}) and torsion vanishes (\ref{T0}). These two conditions do not necessarily follow from the symmetry breaking, but are consistent.
\item Within the subsector defined by (\ref{ee}) and (\ref{T0}), the action is (\ref{action5}),
and it admits the solution (\ref{desitter}). In other words, (\ref{desitter}) is a solution of (\ref{fe1}-\ref{fe3}), thus the ansatze (\ref{hodge},\ref{E},\ref{ee},\ref{T0}) are consistent.
\item Finally, if one looks at the physics around the solution (\ref{desitter}) with the conditions (\ref{hodge},\ref{E},\ref{ee},\ref{T0}) still satisfied, we claim that it is described by the action (\ref{action5}) and thus match Yang-Mills, modified gravity, and Higgs.
\end{enumerate}
The succinct, polynomial form of the action (\ref{action1}) before symmetry breaking stands in contrast to the usual form of the coupled Einstein-Yang-Mills-Higgs action, which requires the inverse and determinant of a metric. To relate the two actions, the metric has become a symmetry breaking field that emerges only for a subset of solutions of the theory. In unifying the gauge group, $\mfr{g}^{spacetime}$, of spacetime with the gauge group, $\mfr{g}^{YM}$, of Yang-Mills fields, we find a unified gauge and spacetime connection which, as a consequence of unification, also includes a metric and Higgs field. The notion that a unification of forces yields a unification of the Higgs field and the metric, with both playing a role in  symmetry breaking, is not new \cite{Percacci}, but it is cleanly realized here.

Much remains to be done to investigate this theory.  The gravitational sector needs to be better understood \cite{Alexandrov,noi1,noi2}.  Since the action and equations of motion are low order polynomials, we believe that progress can be made on the quantization of the unified theory, but this remains to be investigated. In addition, one can consider more general versions of an extended Plebanski action in which $\Phi^3$, in (\ref{action1}), is replaced by a scalar function, $U(\Phi)$, as in \cite{Krasnov}.  While much remains to be done, it is now clear that the line of thought that began with the work of Plebanski and Ashtekar yields a natural and simple proposal for the unification of all known interactions.  

\section*{Acknowledgements}

This research was supported in part by a grant from The Foundational Questions Institute. Research at Perimeter Institute for Theoretical Physics is supported in part by the Government of Canada through NSERC and by the Province of Ontario through MRI.



\end{document}